\documentclass[aps,preprint,nofootinbib,preprintnumbers,superscriptaddress]{revtex4}
\usepackage{amssymb}
\usepackage{amsmath}
\usepackage{amsfonts}
\usepackage{graphicx}
\usepackage[hang,tight,raggedright]{subfigure}
\usepackage{multirow}
\usepackage{subfigure}
\usepackage{float}
\usepackage{graphics}
\usepackage{slashed}
\usepackage{color}
\usepackage{bm}
\usepackage{braket}
\usepackage{psfrag}
\usepackage{booktabs}
\usepackage{latexsym}
\allowdisplaybreaks

\definecolor{darkred}{rgb}{0.7,0.0,0.0}
\definecolor{darkblue}{rgb}{0.0,0.0,0.9}
\definecolor{darkgreen}{rgb}{0.0,0.5,0.0}
\definecolor{brown}{rgb}{0.0,0.0,0.0}

\begin{document}

\title{Combined analysis of jet substructure for Higgs decay to $b\bar{b}$ in vector boson associated production at the 13 TeV LHC }
\author{Fang Tian}
\email{tianfang@pku.edu.cn} \affiliation{School of Physics and State Key Laboratory of
Nuclear Physics and Technology, Peking University, Beijing 100871,
China}

%\date{\today}

\begin{abstract}
We study the Standard Model Higgs and vector boson associated production with large transvese momentum at the 13 TeV LHC, followed by Higgs decay to bottom quark pair.  Using mass-drop tagging and filtering techiques, we obtained the cross section of signal and background. The background mainly come from $Vb\bar{b}$, $t\bar{t}$ and single top production. In order to suppress them further, we combined the mass-drop tagging and filtering analysis with $N$-subjettiness jet shape. After performing $N$-subjettiness identification, the significance can be enhanced to 3.52$\sigma$ with current integrated luminosity of $13.2 {\rm fb}^{-1}$. $H\to b\bar{b}$ decay channel is expected to reach 5$\sigma$ with  $27\,{\rm fb}^{-1}$ at the 13 TeV LHC.
\end{abstract}

%\pacs{12.38.Bx, 12.38.Cy, 12.60.-i}

\maketitle

\section{Introduction}\label{sec:intro}
Since a Higgs boson was discovered with mass $m_H$ near 125 GeV in 2012~\cite{Aad:2012tfa,Chatrchyan:2012xdj}, the measurements of various production and decay modes of the Higgs boson have been performed by both ATLAS and CMS collaborations. After the $\gamma\gamma$ and $ZZ$ decay channels have been measured precisely, the measurements of $WW$ and $\tau^+\tau^-$ decay channels have also reached 5$\sigma$ significance at Run 1 \cite{ATLAS:2014aga,Chatrchyan:2013iaa,Khachatryan:2016vau}, which are crucial to test the electroweak symmetry breaking mechanism. Though $H\to b\bar{b}$ channel has the dominant branching ratio of about 58\% predicted by the Standard Model (SM), searching for $H\to b {\bar b}$ at the LHC is very difficult due to the overwhelmingly large QCD background. It is expected that $H\to b {\bar b}$ can be measured with 5$\sigma$ soon after  several hundreds ${\rm fb}^{-1}$ at 14 TeV\cite{ATLAS:2014php}. Though the cross section of $pp\to HV (V=W\,{\rm or}\,Z)$ is more than an order of magnitude lower than the one of gluon fusion process, this production mechanism is the most promising because the background can be reduced effectively by using the following leptonic decay of $W$ or $Z$. 
At the LHC run 1, both ATLAS and CMS have performed a search for the SM Higgs boson decaying to $b{\bar b}$ in assoiciation with a mass vector gauge boson $W/Z$ \cite{TheATLAScollaboration:2013lia,Chatrchyan:2013zna}. Recently, a simliar research has also been done by ATLAS at the LHC run 2\cite{ATLAS:2016pkl}. The experimentists collected data with an integrated luminosity of 13.2 ${\rm fb}^{-1}$ at a center-of-mass energy of 13 TeV. An excess was observed with significance of 0.42 standard deviations compared with an expectation of 1.94 standard deviations.  In these experiments, the events are selected by requiring two $b$ tagged jets with invariant mass about 125 GeV, which will suffer from large $t\bar{t}$ and multijet QCD background. 

A promising method to suppress the background is to investigate Higgs boson production with high transverse momentum, i.e. $p_{T,H}\sim200$\,GeV or more. In this region the hadronic decay products of Higgs will be clustered into a single fat jet, and jet substructure techniques are needed to identify it. Ref.~\cite{Butterworth:2008iy} has proposed a method to identify highly boosted Higgs with hadronic decay, which can be divided to two steps: mass-drop tagging and filtering. With this method, the underlying events and multijet QCD background can be suppressed signifcantly, and make it possible to measure $H\to b\bar{b}$ with a high significance under the current luminosity. An update study was performed at 14 TeV \cite{Butterworth:2015bya}. 

Though mass-drop tagging and filtering are very successful, it is still very interesting to develop more jet substructure techniques to explore the hadronic decay of Higgs boson. First, it is necessary to suppress multijet and $t\bar{t}$ background further. This is not only helpful to make the measurement of $H\to b\bar{b}$ more precise, but also helpful for many other researches involving mass-drop and filtering, i.e. Higgs pair production with $H\to b\bar{b}$ \cite{deLima:2014dta}. Second, it can help us to study the Higgs boson decay to light jets, understanding the Higgs Yukawa coupling to light quarks ($u$, $d$, $c$, $s$)\cite{Perez:2015lra,Soreq:2016rae,Gao:2016jcm,Yu:2016rvv,Carpenter:2016mwd}.  During the past few years, a lot of jet substructure techniques have been developed \cite{Almeida:2008yp,Ellis:2009su,Ellis:2009me,Krohn:2009th,Gallicchio:2010dq,Thaler:2010tr,Stewart:2015waa,Thaler:2015xaa}. 

 In this paper, we explore the SM Higgs boson produced in association with a $W$ or $Z$ boson and decaying to $b\bar{b}$ pair at the 13 TeV LHC. After doing mass-drop and filtering, we find that the Higgs jet still has other feature different from QCD jet and top jet, and QCD background can be eliminated further by using $N$-subjettiness variable. 
 
The paper is organized as follows. In section~\ref{sec:method}, we describe the two different jet substructure methods and make a comparison between them.  Section~\ref{sec:eventsel} describes the detailed kinematic cuts. In section~\ref{sec:numres}, we calculate the signal and background distribution and make a discussion. We conclude in section~\ref{sec:conc}
 
\section{Method}\label{sec:method}
When a heavy resonance with transverse momentum $p_T$ much greater than its mass $M$, its hadronic decay products are almost collinear and may be recombined into a single fat jet by jet algorithms. For two-body decay,the radius of the fat jet can be expressed roughly as
\begin{align}
	R_{jj}  \simeq \frac{M}{\sqrt{z(1-z)}\,p_T}\,,
\end{align}
where $z$ is mometum fraction of decay product. In ref.~\cite{Butterworth:2008iy}, the authors proposed a method to reconstruct highly boosted Higgs boson in $b\bar{b}$ final state, which can be divided to two steps: mass-drop tagging and filtering. For a fat jet obtained from Cambridge/Aachen jet algorithm\cite{Dokshitzer:1997in,Wobisch:1998wt} with radius $R$ , the procedure of mass-drop is as follow
\begin{itemize}
	\item[(1)] Undoing the last step of jet clustering, splitting the fat jet $j$ into two subjets, denoted by $j_1$ and $j_2$ with $m_{j_1}>m_{j_2}$.
    \item[(2)] Introduce two dimensionless parameters $\mu$ and $y_{\rm cut}$. Check to see whether the following two conditions are satisfied: $m_{j_1}< \mu m_{j}$ and $y=\frac{{\rm min}(p_{T,j1}^2,p_{T,j2}^2)}{m_J^2}\Delta R_{j_1,j_2}^2 > y_{\rm cut}$, mean that there is a significant mass drop from the daughter jets to mother jet, and no large asymmetry in the spliting, respectively.
     \item[(3)] If the above two conditions are satisfied,then tag the original jet $j$ as a Higgs candidate and exit the iteration.  Otherwise, discard the softer subjet $j_2$ and go back to step 1 with $j_1\to j$. 
\end{itemize}
Jet formed by hadronic decay of Higgs boson is different from QCD jet. A Higgs jet consist of two narrow hard $b$ jets and some soft jets, while the radiations in a QCD jet distribute more uniformly. So after jet splitting,  mass-drop will happen for Higgs jet but not for QCD jet.  

Mass-drop method can distinguish the fat jet formed by boosted heavy resonance from QCD jet, but it is still not easy to reconstruct the the Higgs boson mass at the transverse momentum region $p_T\sim 200$\,GeV. Because the jet radius $R_{b\bar{b}}\sim 2m_H/p_T$ is quite large, there will be a lot of radiations from underlying events in the jet cone, which is not relevant to the hadronic decay of Higgs boson and must be eliminated. To achieve this, the authors in ref.~\cite{Butterworth:2008iy} propose another method to filter the unrelevant radiation in Higgs jet. They introduced an angular scale $R_{\rm filt}$, which denotes the angular separation of two $b$ jets from the Higgs decay. Then the fat jet can be resolved at the filtering scale $R_{\rm filt}$, and only three hardest subjets are kept.  In practice, the optimal set of filter scale is $R_{\rm filt}={\rm min}(R_{b\bar{b}}/2,\,0.3)$. 

In ref.~\cite{Butterworth:2008iy}, we can see that the signal/background ratio is enhanced significantly after mass-drop and filtering procedures. The donminant backgrounds come from $pp\to V+b+{\bar b}$ and $pp\to t + \bar{t}$ processes.These backgrounds would be larger when the centre-of-mass energy of the LHC increases. If we want to measure the Higgs coupling to vector gauge boson and to $b\bar{b}$ more precisely, we need to go a step further to suppress the background. To achieve this, in the following context we will try to combine the above methods with another jet substructure techique.

In ref.~\cite{Thaler:2010tr}, the authors introduced a jet shape - $N$-subjettiness, which can identify a boosted heavy particle with $N$-prong hadronic decay products effectively, such as two-prong Higgs or W/Z boson and three-prong top quark. Given $N$ axes $\hat{n}_k$, $N$-subjettiness jet shape is defined as
\begin{align}\label{eq:defNjettiness}
\tau_N=\frac{\sum_{i\in J }p_{T,i}{\rm min}(\Delta R_{ik})}{\sum_{i\in J }p_{T,i} R}\,,
\end{align}
where $R$ is radius of the jet, and $\Delta R_{ik}$ is angular distance between particle $i$ and axe $\hat{n}_k$. The smaller $\tau_N$ is, the closer to the $N$ axes the radiation in jet would be. Broadly, $\tau_N$ of a jet would decrease when $N$ increases, because some of particles will meet a closer axe when new axes are added in the jet. But there is a significant difference between $N$-prong jet and QCD jet. For a boosted Higgs boson decaying to $b\bar{b}$, most of the particles in the fat jet are clustered around the two directions of $b$ quark pair, so its $\tau_2$ would be much smaller than $\tau_1$. While for a QCD jet, because the particles distribute uniformly in the jet, the difference between $\tau_{2}$ and $\tau_{1}$ is not as significant as the case for Higgs boson. Thus a discriminating variable can be introduced
\begin{align}\label{eq:tauij}
\tau_{N,N-1}=\frac{\tau_N}{\tau_{N-1}}\,,
\end{align}
to find an $N$-prong candidates.

\begin{figure}
	\begin{center}
		\subfigure[Higgs jet]{
			\label{fig:jsub1}\includegraphics[width=0.26\textwidth]{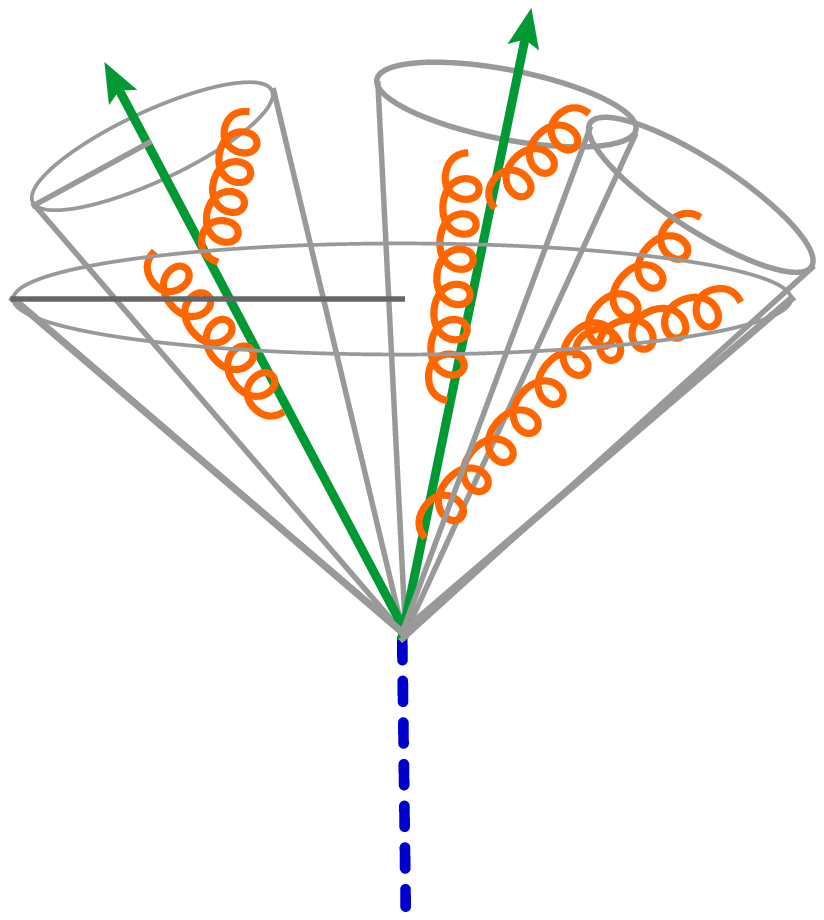}
		}
		\subfigure[gluon jet]{
			\label{fig:jsub2}\includegraphics[width=0.26\textwidth]{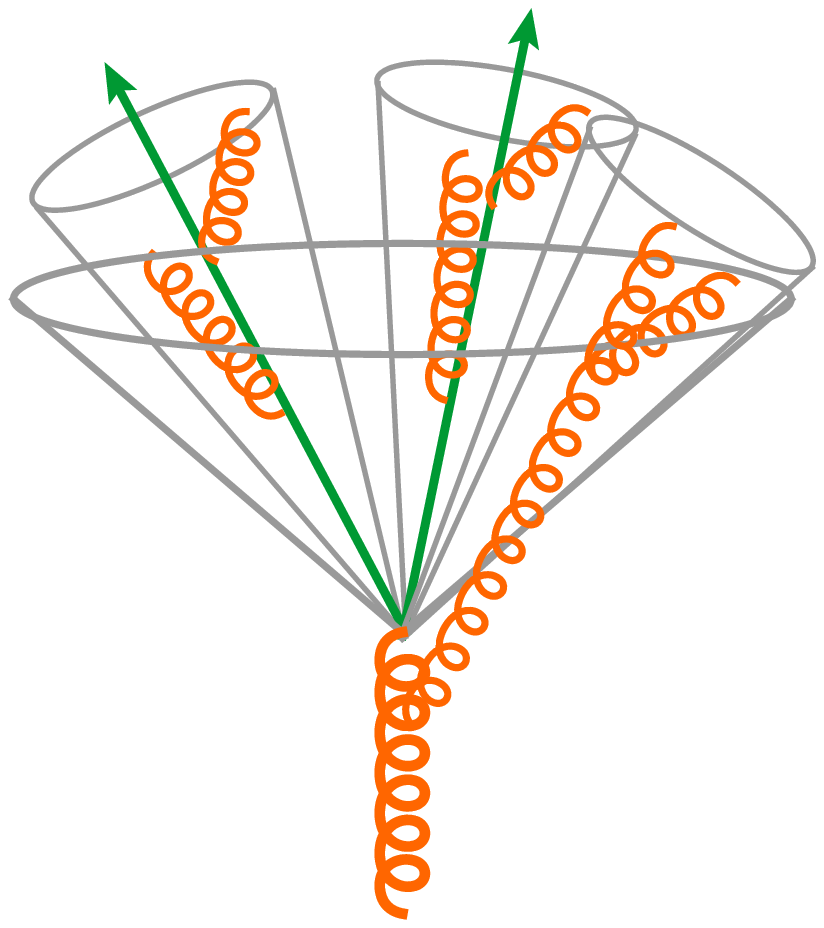}
		}
		\subfigure[top jet]{
			\label{fig:jsub3}\includegraphics[width=0.26\textwidth]{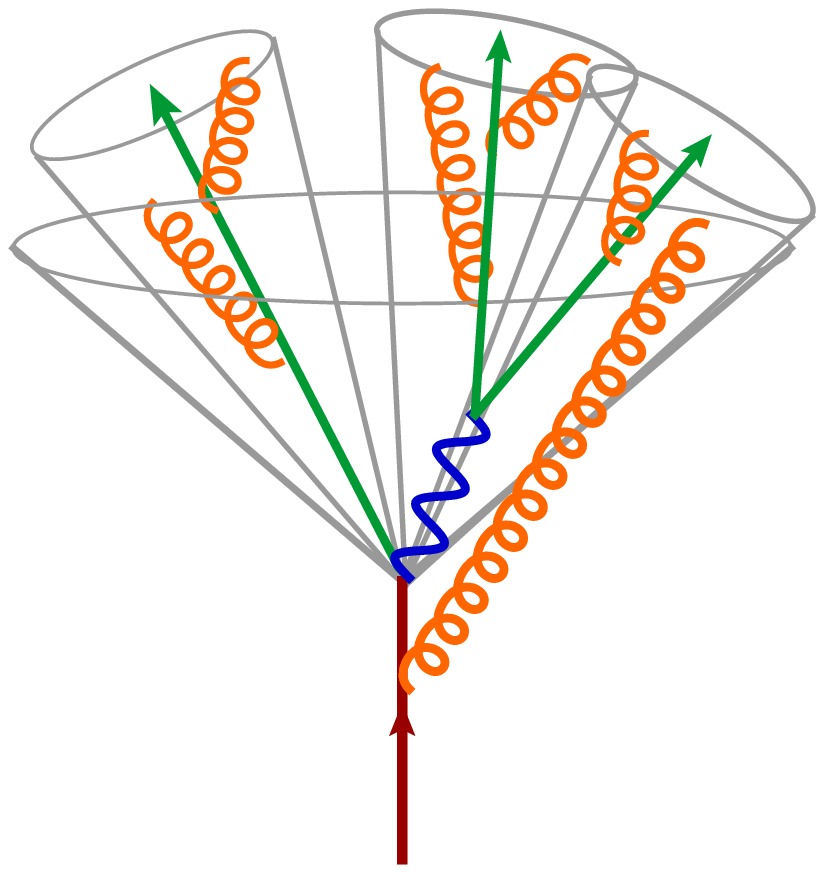}
		}
	\end{center}
	\vspace{-0.5cm}
	\caption{\label{fig:jsub} The jet substructure of Higgs, gluon and top jet.
	}
\end{figure}

The next question need to be answered is that whether $N$-subjettiness identification is still necessary after mass-drop tagging and filtering procedures have been performed to a fat jet. Our answer is yes. In order to reconstruct the Higgs boson mass precisely, three hardest subjets are kept to include the ${\cal O}(\alpha_s)$ radiation. For a Higss jet, the three subjets dominantly consist of two hard $b$ subjets and a soft gluon subjet, as shown in fig.~\ref{fig:jsub1}. Its structure is different for QCD jet and top jet. For the background process $pp\to V+b+\bar{b}$, most $b\bar{b}$ final states come from the gluon splitting. Unlike a color-singlet Higgs, gluon is color-octet and can take part in stronge interaction. As shown in fig.~\ref{fig:jsub2}, the gluon subjet in a gluon jet can radiate from the off-shell hard gluon and it can take a large part of enery of its off-shell mother gluon. This means that for a gluon jet the energy distribution is uniform among the three selected subjets, while for a Higgs jet the two $b$ subjets would be dominant. As a result, $\tau_{21}$ of Higgs jet would be smaller than the one of gluon jet even after mass-drop tagging and filtering. For the background process $pp\to t\bar{t}$ and single  top production, because top quark can take part in the strong interaction, the fat jet could consist of two $b$ subjets and a gluon radiated from top quark. This gluon does not strongly correlate with $b$ quark and its enery could be competitive to the two $b$ quark. Furthermore, a boosted top quark with hadronic decay can be contamination if the charm jet from $W$ decay is mis-tagged as a $b$-jet. In this case the fat jet formed by hadronic top decay would be three-prong, which has a large $\tau_{2}$, just as shown in fig.~\ref{fig:jsub3}. So $\tau_{21}$ of the top jet in $t\bar{t}$ and single top production process would be larger than the one of Higgs jet. Now, we can expect that the $Vb\bar{b}$ and $t\bar{t}$ background would be suppressed further with $N$-subjettiness discrimination. 

\section{Event selection}\label{sec:eventsel}
The events are generated using the LO mode of \texttt{MadGraph5\_aMC@NLO}\cite{Alwall:2014hca}  followed by \texttt{PYTHIA6}\cite{Sjostrand:2006za} parton shower generator. The MSTW2008nlo\cite{Martin:2009iq} parton distribution function sets are used. Both signal and background samples are generated with 0/1/2 jet parton level matching, based on the default $k_T$-jet MLM scheme in \texttt{MadGraph5\_aMC@NLO}.  Jet algorithm and substructure techiques are performed with \texttt{FASTJET} package\cite{Cacciari:2011ma}.

We focus on three categories of events, which contain 0, 1 and 2 electrons (or muons) in the final state, targeting  the $Z\to \nu \nu$, $W\to l \nu$ and $Z\to \nu \nu$ decay modes respectively. Broadly, we select the events where a high transverse momentum fat jet containg two highly boosted b-jets is reconstructed together with 0, 1 or 2 charged leptons (electrons or muons). Cambridge/Aachen algorithms are performed with $R=1.2$ to find fat jet, and then mass-drop tagging and filtering are used to reconstrut the Higgs boson four momentum. We select the leading jet as Higgs candidate with transverse momentum $p_T>200$\,GeV and rapidity $|y|<3.5$.   The mass drop threshold and asymmetry requirement are chosen as $\mu=0.67$ and $y_{\rm cut}=0.09$, just as ref.~\cite{Butterworth:2008iy}. If there is a significant mass drop in the fat jet, we perform filtering on the jet's constituents with $R_{\rm filter}={\rm min}\{R_{b\bar{b}}/2,\,0.3\}$. We take the three leading filtered jets to reconstruct the Higgs jets. To suppress the background contamination,  we apply a mass window $110<m_{b\bar{b}}<130$\,GeV  on the invariant mass of Higgs boson candidate. The jet is tagged as a Higgs jet only if there are two filtered jets in the three satisfying the $b$-tagging criteria with efficiency 70\%. The $c$-to-$b$ and light-jet-to-$b$ mis-tagging probabilities are assumed of 10\% and 1\%, respectively\cite{deLima:2014dta}.  In order to suppress the background from $t{\bar t}$ and single top production, we also reject the events containing additional  jets with $|\eta|<2.5$ and $p_{T}>30$ GeV. 

For the three categories of events corresponding to the three leptonic decay modes of vector boson, the analysis consists of the following cuts:
\begin{itemize}
	\item[(a)]0-lepton: Missing transverse momentum $E_{T,{\rm miss}}>200$\,GeV\,;
	\item[(b)]1-lepton: Missing transverse momentum $E_{T,{\rm miss}}>30$\,GeV plus a lepton $l$ ($l=e\,{\rm or}\,\mu$)  with $p_l>30$\,GeV and $|\eta_l|<2.5$, consistent with a $W$ of nominal mass with $p_T>200$\,GeV\,;
	\item[(c)]2-lepton: An $l^+l^-$ ($l=e \,{\rm or}\,\mu$) pair with an invariant mass $80<m_{ll}<100$\,GeV and $p_{T,ll}>200$\,GeV, each lepton $l$ satisfys $p_l>10$\,GeV and $|\eta_l|<2.5$\,.	
\end{itemize}
In cases (a) and (b), the selected leptons should be isolated, having $\sum_i p_{T,i}$ less than  10\% of it transverse momentum within a cone of $\Delta R = 0.3$ around it.  
%To suppress the background, the events containing leptons with $|\eta_l|<2.5$ and $p_{T,l}>30$\,GeV apart from those used to reconstruct the leptonic vector boson should be rejected.  
To suppress $t\bar{t}$ and single top backgrounds, we reject the events with $E_{{\rm miss},T}>20$\,GeV for 2-lepton case. We also veto additional isolated leptons with $|\eta_l|<2.5$ and $p_{T,l}>30$ for 0-lepton and 1-lepton modes. In the following,  the above kinematic cuts and jet mass window$110<m_{b\bar{b}}<130$\,GeV are called as "basic cuts".

\section{Numerical Results and Discussion}\label{sec:numres}
\begin{figure}[t]
	\begin{center}
		\subfigure[0-lepton]{
			\label{fig:mod2_mbb}\includegraphics[width=0.450\textwidth]{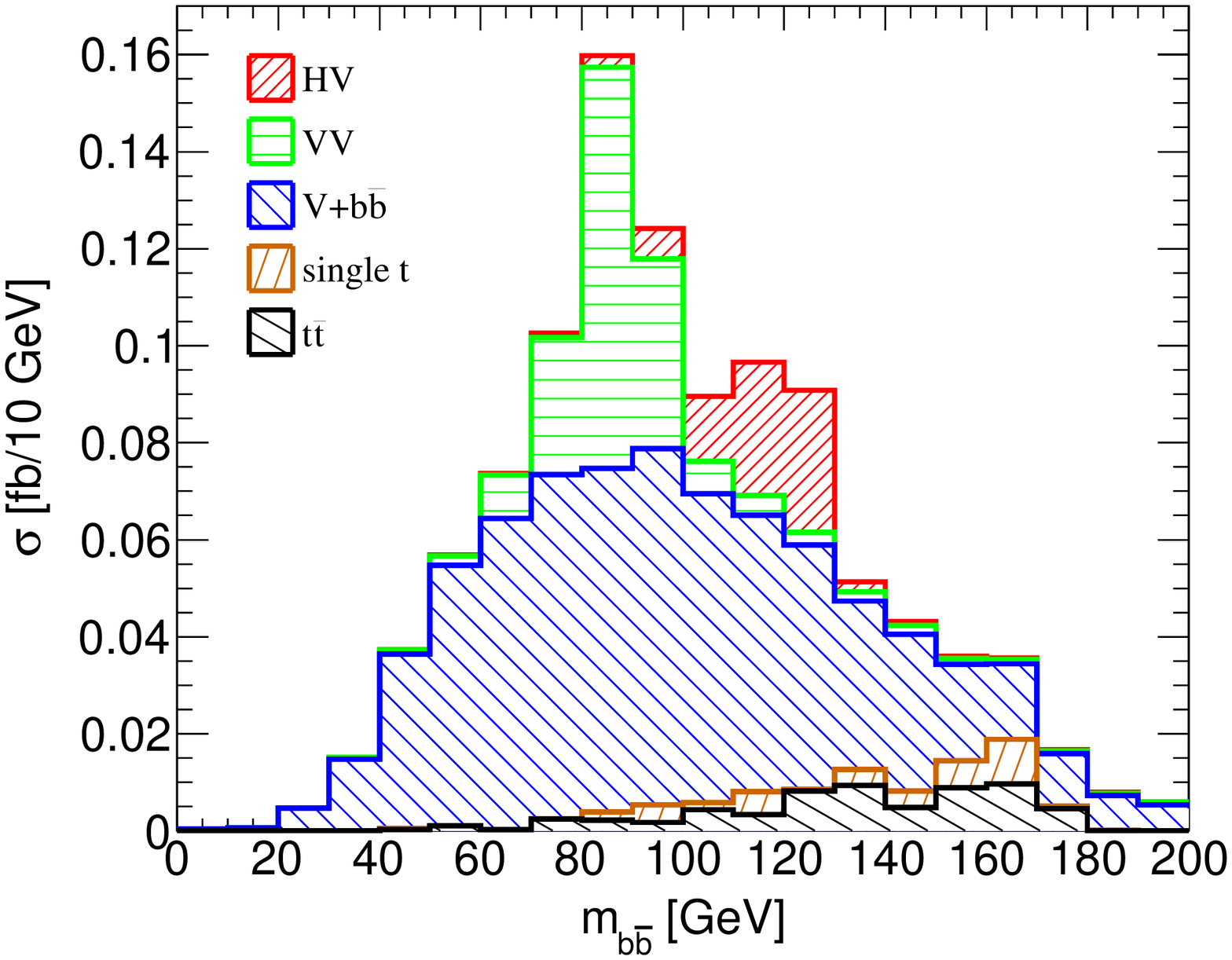}
		}
	    \subfigure[1-lepton]{
					\label{fig:mod3_mbb}\includegraphics[width=0.450\textwidth]{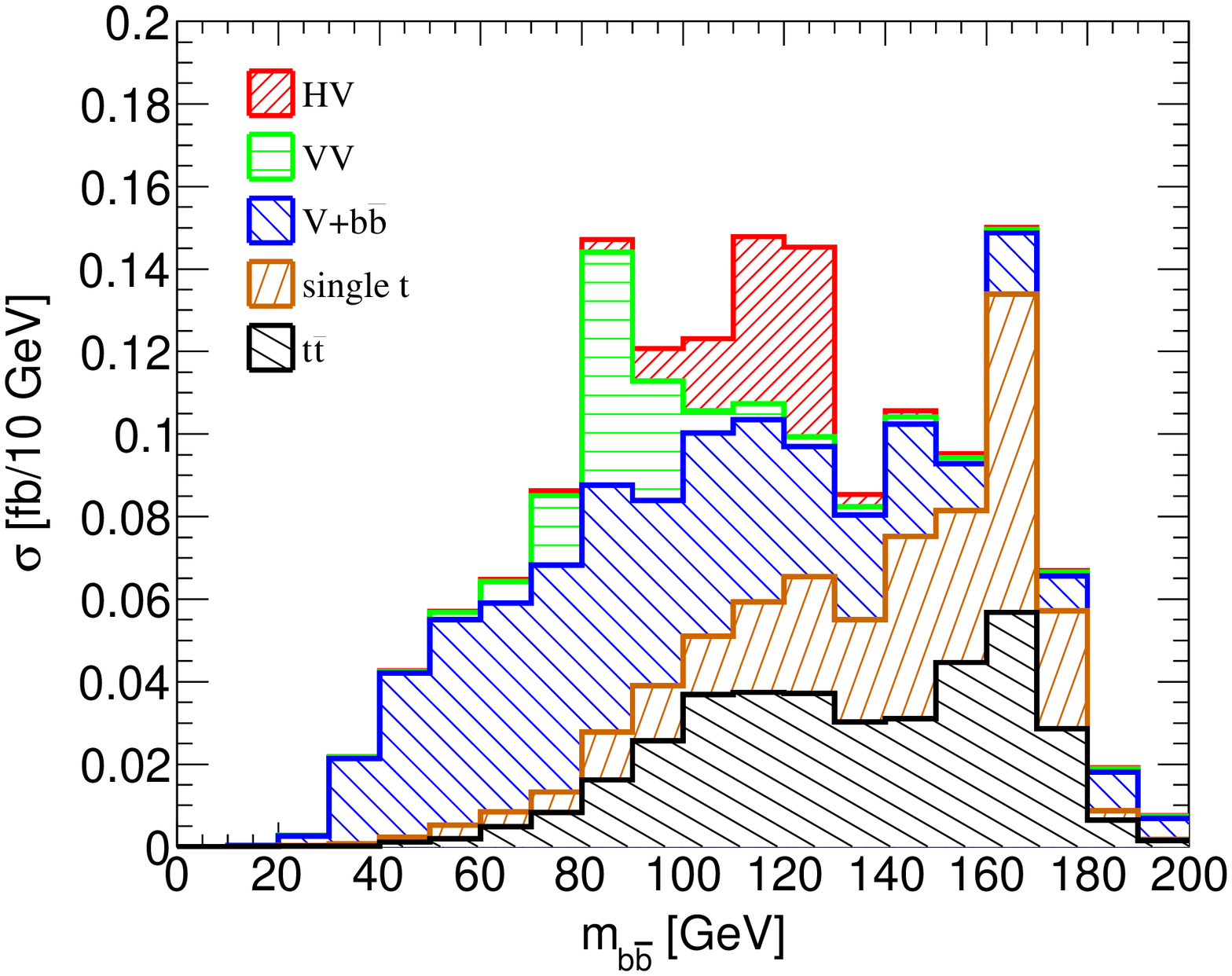}
	     }
	    \\
		\subfigure[2-lepton]{
			\label{fig:mod1_mbb}\includegraphics[width=0.450\textwidth]{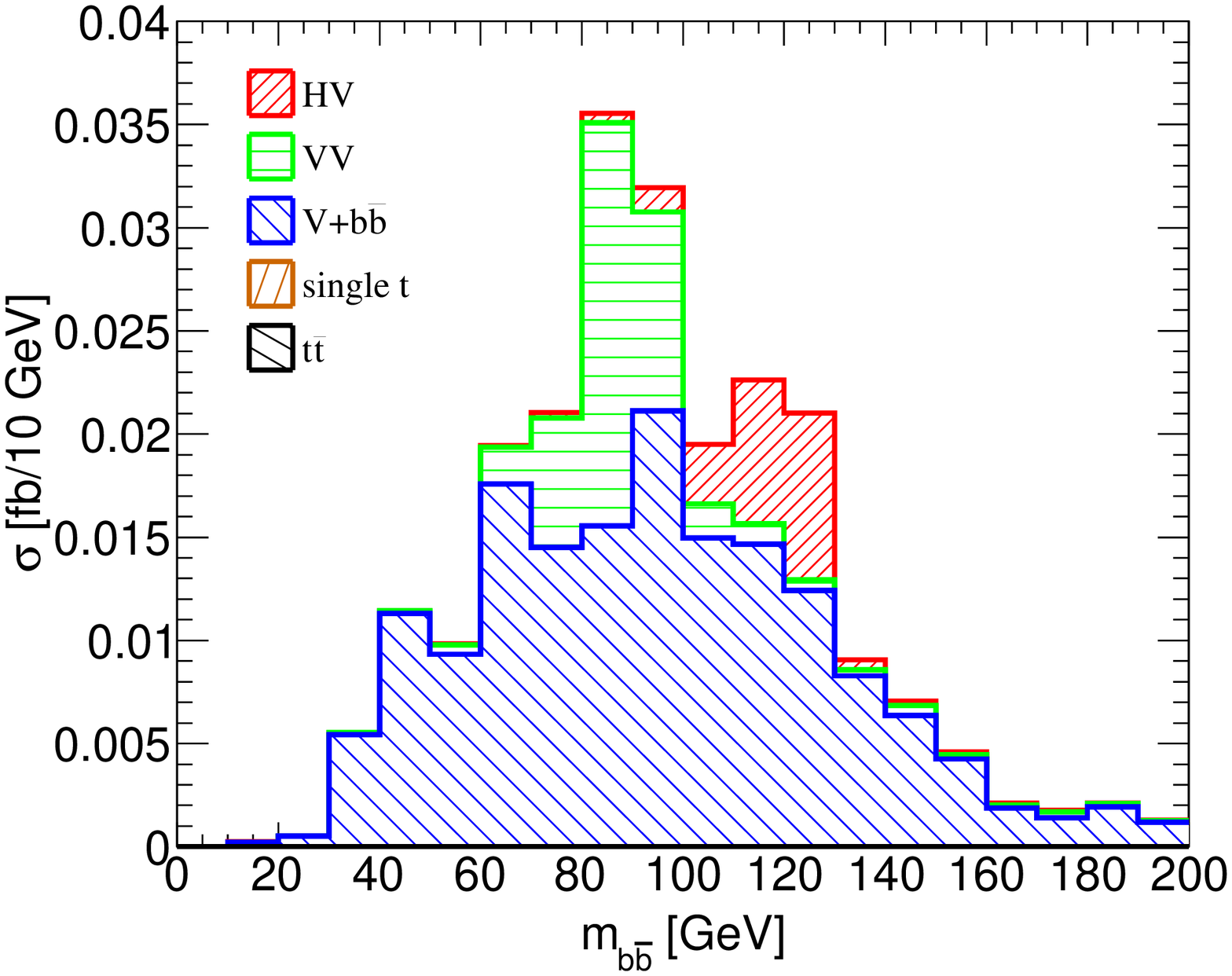}
		}
		\subfigure[Combined]{
			\label{fig:total_mbb}\includegraphics[width=0.450\textwidth]{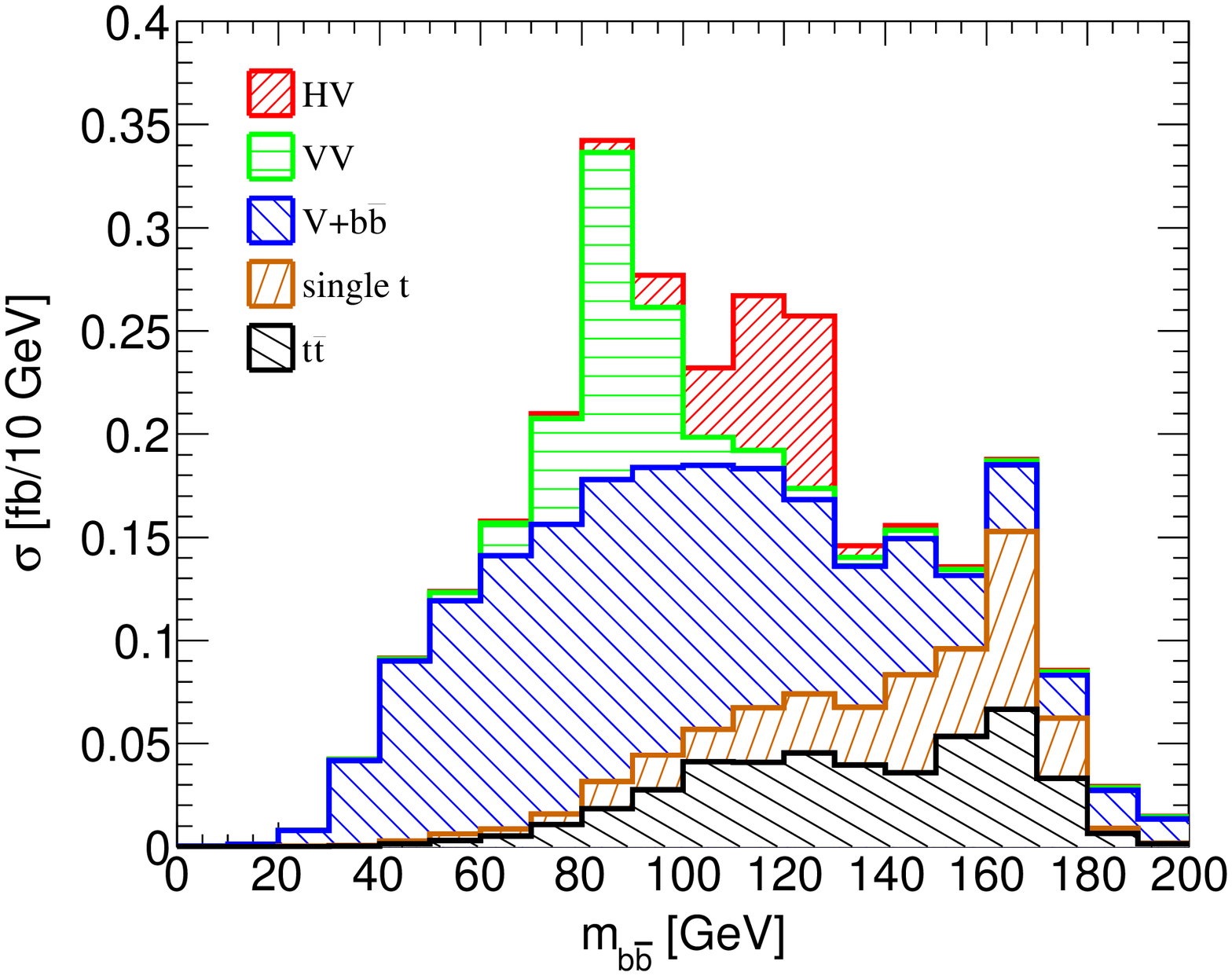}
		}
	\end{center}
	\vspace{-0.5cm}
	\caption{\label{fig:mbb_mdrop} The invariant mass distribution of the filtered jets with mass-drop tagging at the 13\,TeV LHC. 
		}
\end{figure}

In fig.~\ref{fig:mbb_mdrop}, we show the invariant mass distribution of the filtered jets with mass-drop tagging at the 13\,TeV LHC for 0-lepton, 1-lepton, 2-lepton cases individually and their combination. The single top background consists of $t-$channel, $s-$channel single top production and associated production of $t\,W$. The distributions of vector gauge boson pair ($VV$, $V=W$or $Z$) background peak at $90$ GeV due to $Z$ decay to $b\bar{b}$, which can be clearly distinguished from the spectrum peaking at $m_H=125$ GeV.  So the contribution from $pp\to VV$  is highly suppressed. For 0-lepton and 2-lepton modes, almost all of the backgrounds come from $pp\to Z(\to  l+\bar{l})+b+\bar{b}$. The $t\bar{t}$ and single top backgrounds are highly suppressed because of the additional lepton and jet veto. In addition, the fully leptonic decay of $t\bar{t}$ and $tW$ contribute zero to the background due to the limit of total missing energy $E_{{\rm miss},T}<20$\,GeV for 2-lepton case. However, we should treat $t\bar{t}$ and single top background seriously for 1-lepton case. It is because that the final states of $t\bar{t}$ or $t\,W$ process can consist of a hadronic decay (anti-)top $t\to W(\to c+s) +b$ and a recoiling $W\to l+\nu_l$. At large transverse momentum region of top quark,   $c$ and $b$ jets would be clustered into a fat jet, which can contaminate to the Higgs-like jet with $c$-to-$b$ mis-tagging. Furthermore,  $t-$channel and $s-$channel single top background also play a non-negiliable role, because at leading order the final states consist of a $W$ boson and two jets, one of which is $b$ jet from top decay.
%The $pp\to t(\to W^+(l^+ +\nu_l)+b)+\bar{t}(\to W^-(l^- + \bar{\nu}_l)+\bar{b})$ 

\begin{table}
	\begin{center}
		\begin{tabular}{c|c|c|c|c}
			\hline
			\hline
		    [fb]  &~~~  0-lepton ~~~& ~~~ 1-lepton ~~~ & ~~~ 2-lepton ~~~ & ~~~Combined~~~
			\\
			\hline					
			$HV$              &  0.5669   &  0.8641  &  0.1504  &  1.581
			\\        	
%			\hline
			$t\bar{t}$        &   0.11536  & 0.7469  &  0         &   0.8623
			\\
%			\hline
			single top        &  0.05047  &  0.501    &  0        &   0.5515 
			\\
%	    	\hline
            $V+b\bar{b}$  &   1.074     &  0.7573   & 0.2708  &  2.102  
            \\
%			\hline
            $VV$              & 0.06648  &  0.0612  & 0.01514 &   0.1428
            \\
%	    	\hline
			total b.g.        &  1.306      &   2.066   &  0.286  &    3.659
			\\
			\hline
			$S/B$             & 0.434      &  0.4182   & 0.5261  &  0.4322
			\\
			\hline
			$S/\sqrt{B}$ [13.2 ${\rm fb}^{-1}$]&  1.8  &  2.18 & 1.02  & 3.0
			\\
			$S/\sqrt{B}$ [30 ${\rm fb}^{-1}$]  &  2.72  & 3.29  &  1.54 & 4.53
			\\
			$S/\sqrt{B}$ [100 ${\rm fb}^{-1}$]  &  4.96  & 6.01  &  2.81 & 8.27
			\\
			\hline
			\hline
		\end{tabular}
	\end{center}
	\caption{\label{tab:sb1}Signal and background cross section with basic kinematic cuts. Mass-drop and filtering are performed for leading jets. The mass window of the filtered jet with two $b$-tagged subjets is $110<m_{b{\bar b}}<130$ GeV. }
\end{table}

Table~\ref{tab:sb1} shows the signal and background cross section with basic kinematic cuts. Mass-drop tagging and filtering have been performed for leading jets. The mass window of the reconstructed fat jet is $110<m_{b{\bar b}}<130$ GeV. Comparing with the other two cases, 1-lepton analysis gives the most sigificant $S/\sqrt{B}$, because of the largest signal cross section. Combining the three channels,  the signal of $H\to b\bar{b}$ can be seen at a significance of  3$\sigma$ with the current integrated luminosity of $13.2 {\rm fb}^{-1}$ \cite{ATLAS:2016pkl}. The siginificance can be expected to reach 8.27$\sigma$ at the further integrated luminosity of $100 {\rm fb}^{-1}$.

\begin{figure}[t]
	\begin{center}
		\subfigure[]{
			\label{fig:tau21_Vbb}\includegraphics[width=0.4\textwidth]{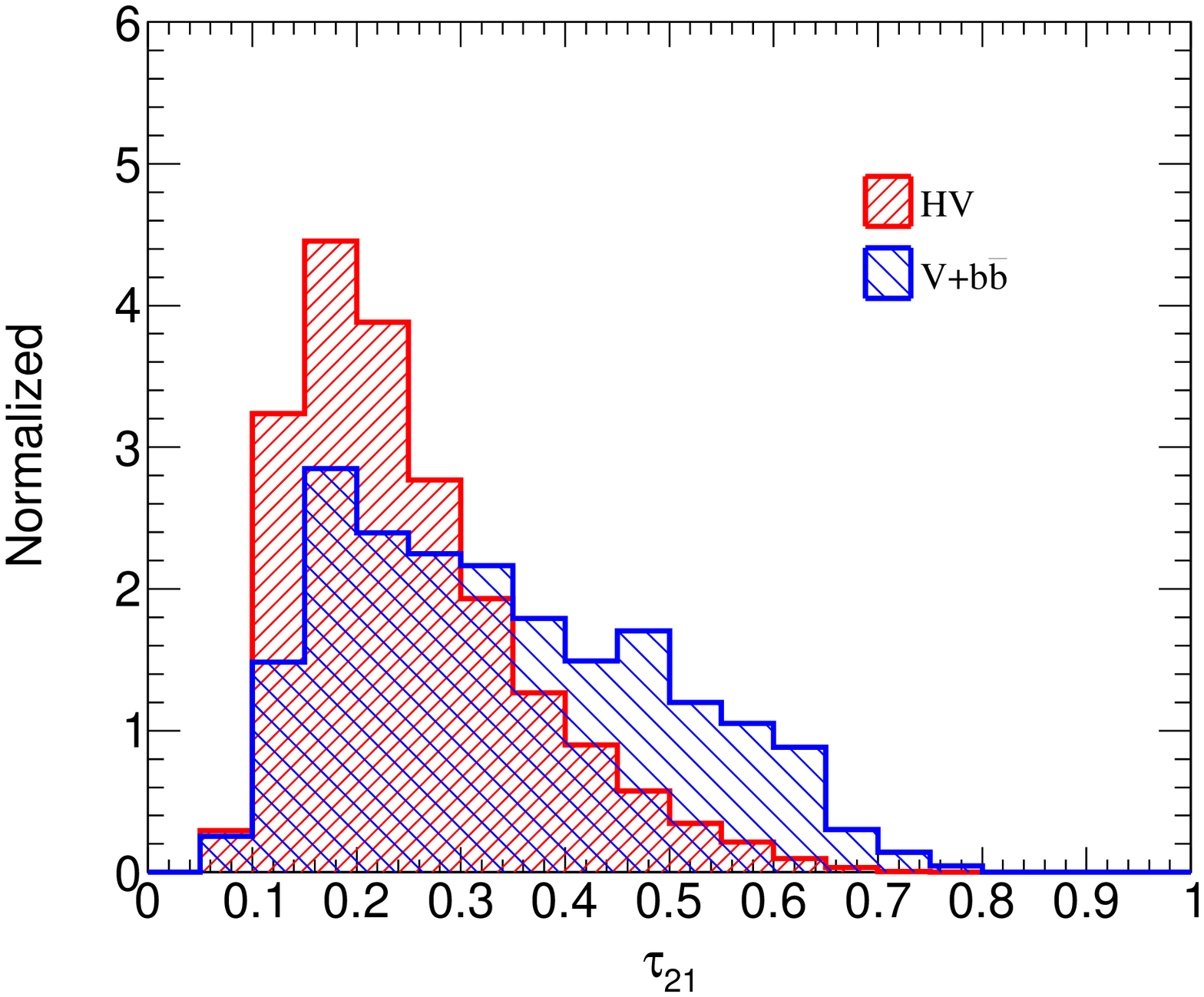}
		}
		\subfigure[]{
			\label{fig:tau21_ttx}\includegraphics[width=0.4\textwidth]{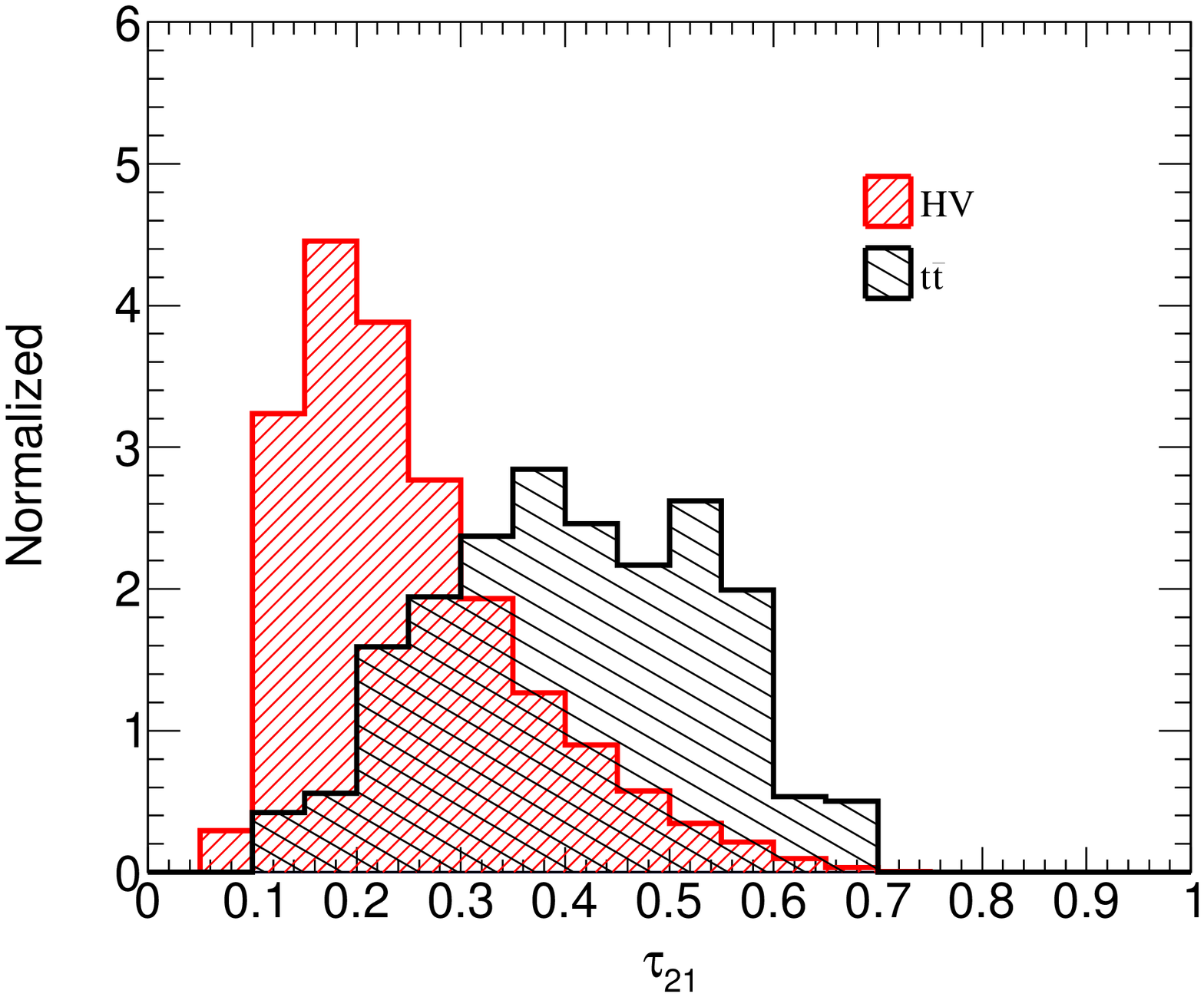}
		}\\
		\subfigure[]{
			\label{fig:tau21_1t}\includegraphics[width=0.4\textwidth]{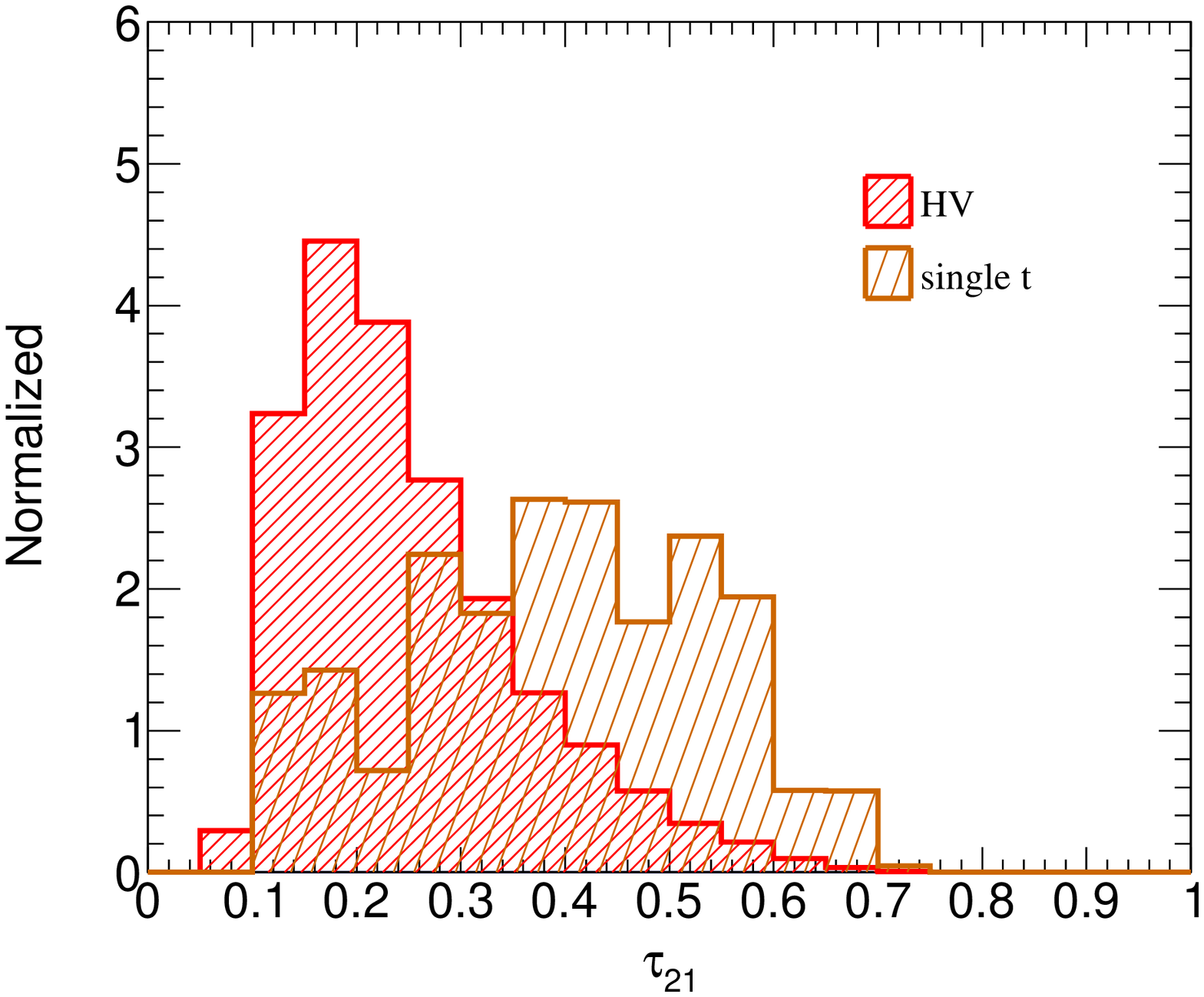}
		}
		
	\end{center}
	\vspace{-0.5cm}
	\caption{\label{fig:tau21_dis} $\tau_{21}$ distribution
	}
\end{figure}

\begin{table}
	\begin{center}
		\begin{tabular}{c|c|c|c|c}
			\hline
			\hline
			[fb]  &~~~  0-lepton ~~~& ~~~ 1-lepton ~~~ & ~~~ 2-lepton ~~~ & ~~~Combined~~~
			\\
			\hline					
			$HV$              &  0.4149 & 0.6324 & 0.1101 & 1.157
			\\        	
			%			\hline
			$t\bar{t}$        &  0.02601 & 0.1684 & 0 & 0.1944  
			\\
			%			\hline
			single top        &  0.01425 & 0.1415 & 0 & 0.1557 
			\\
			%	    	\hline
			$V+b\bar{b}$  &  0.4956 & 0.3494 & 0.125 & 0.97    
			\\
			%			\hline
			$VV$              &  0.04865 & 0.04479 & 0.01108 & 0.1045
			\\
			%	    	\hline
			total b.g.        &   0.5845 & 0.7041 & 0.136 & 1.425
			\\
			\hline
			$S/B$             &  0.7099  & 0.8983 & 0.8094 & 0.8125
			\\
			\hline
			$S/\sqrt{B}$ [13.2 ${\rm fb}^{-1}$]& 1.97  & 2.74 & 1.08 & 3.52
			\\
			$S/\sqrt{B}$ [30 ${\rm fb}^{-1}$]  & 2.97 & 4.13 & 1.64 & 5.31
			\\
			$S/\sqrt{B}$ [100 ${\rm fb}^{-1}$] & 5.43 & 7.54 & 2.99 & 9.7
			\\
			\hline
			\hline
		\end{tabular}
	\end{center}
	\caption{\label{tab:sb2} Combined the results of Tab.~\ref{tab:sb1} with $N$-subjettiness jet shape cut $\tau_{21}<0.3$ on the mass-drop tagged fat jet. }
\end{table}

Next, we combine the above analysis with $N$-subjettiness indentification. Fig.~\ref{fig:tau21_dis} shows the comparison of $\tau_{21}$ distribution of mass-drop tagged  jets between signal and different background, where $\tau_{21}$ is defined as eq.~(\ref{eq:tauij}). The $\tau_{21}$ distributions of $Vb\bar{b}$, $t\bar{t}$ and single top background are wider than the one of signal $HV$. In addition,  $t\bar{t}$ and single top processes have larger average value of $\tau_{21}$ than the one of $Vb\bar{b}$. This is because that the fat jets in $t\bar{t}$ and single top production can be formed by boosted top quark with hadronic decay, which consist of three separated hard subjets and have large $\tau_2$ value. 
In order to suppress the background from $Vb\bar{b}$, $t\bar{t}$ and single top production further, we can perform an additional $\tau_{21}$ cut on the selected events. An optimum choice of $\tau_{21,{\rm cut}}$  should reject more background events but keep more signal events. Here we choose $\tau_{21,{\rm cut}}=0.3$. Tab.~\ref{tab:sb2} presents the signal and background cross section and the significance after doing $N$-jettiness identification. For the combined analysis, though the signal cross section decreases about 27\% , $Vb\bar{b}$, $t\bar{t}$ and single top background decrease about 54\%, 77\%, and 73\%, respectively. The significance can be enhanced to 3.52$\sigma$ with the current integrated luminosity of $13.2 {\rm fb}^{-1}$. With $N$-jettiness identification, $H\to b\bar{b}$ decay channel is expected to reach 5$\sigma$ with  $27 {\rm fb}^{-1}$ at the 13 TeV LHC. 
%Hence, we introduce a ratio $R_\tau$ to perform the $N$-jettiness identification
%\begin{align}
%R_\tau(\tau_{21,{\rm cut}})=\frac{\sigma(\tau_{21}<\tau_{21,{\rm cut}})}{\sigma}\,,
%\end{align}
%where $\sigma$ denotes the cross section with basic kinematic cuts, mass-drop tagging and jet filtering. $\sigma(\tau_{21}<\tau_{21,{\rm cut}})$ denotes the above cross section with additional constraint of $\tau_{21}<\tau_{21,{\rm cut}}$.When we fix $\tau_{21,{\rm cut}}$, the jet with larger $R_\tau$

\section{Conclusion}\label{sec:conc}
In this paper, we studied the SM Higgs and vector boson associated production with large transvese momentum at the 13 TeV LHC, followed by Higgs decay to bottom quark pair. We performed the analysis in 0-lepton, 1-lepton and 2-lepton modes separately, and the then combined them together. Using mass-drop tagging and filtering techiques, we obtained the cross section of signal and background. The sigal can be observed with a significance of 3.0 at current integrated luminosity of 13.2 ${\rm fb}^{-1}$. The background mainly come from $Vb\bar{b}$, $t\bar{t}$ and single top production. In order to sppress them further, we combined the above analysis with $N$-subjettiness jet shape. After performing $N$-subjettiness identification, the significance can be enhanced to 3.52$\sigma$ with $13.2\,{\rm fb}^{-1}$. $H\to b\bar{b}$ decay channel is expected to reach 5$\sigma$ with  $27\,{\rm fb}^{-1}$ at the 13 TeV LHC. 

\begin{acknowledgments}
We would like to thank Ze Long Liu, Jian Wang and Felix Yu for helpful discussions. This work is partially supported in part by the National Natural Science Foundation of China under Grants No.~11475006. 
\end{acknowledgments}
\bibliography{ppHV}
\end{document}